\documentstyle[pra,aps]{revtex}

\title{On finite-size effects in computer simulations using the Ewald
 potential}
\author{Francisco Figueirido, Gabriela S. Del Buono and Ronald M. Levy}
\address{Department of Chemistry\\
Wright-Rieman Laboratories\\
Rutgers, The State University of New Jersey\\
Piscataway, NJ 08855--0939}
\date{\today}

\begin{document}

\bibliographystyle{prsty}

\widetext
\maketitle

\begin{abstract}
We discuss the origin and relevance for computer simulations of a strong
finite-size effect that appears when using the Ewald summation formula.  It
can be understood as arising from a volume-dependent shift of the potential in
a finite, periodic box relative to the infinite volume limit. This shift is
due to the fact that the ``zero of energy'' for a periodic system cannot be
defined by letting the interacting particles be separated by an infinite
distance; the correct definition corresponds to setting its $\bbox k=\bbox 0$
Fourier mode to zero. The implications of this effect for computer simulations
are discussed.
\end{abstract}

\narrowtext

\section{Introduction}
\label{sec:intro}

The correct and efficient treatment of the long-range electrostatic
interactions in simulations of aqueous solutions continues to be an issue of
intense research. The Ewald summation technique
\cite{ewald:21,kornfeld:24,allen:tildesley} appears to be a promising approach
for the consistent treatment of these interactions.  Several groups have
discussed that the truncation of the electrostatic forces, either spherical or
using the ``minimum image'' convention, results in severe modifications of the
dielectric behavior
\cite{neumann:dielectric-sums,neumann:ladd,neumann:steinhauser:ewald,%
mahfoud:ewald,alper:levy:reaction-field}.
There are indications that these modifications have quite noticeable effects
on the behavior of aqueous solutions in computer simulations
\cite{schreiber:steinhauser:one,%
schreiber:steinhauser:two,schreiber:steinhauser:three,%
bader:chandler,delbuono:94}.

In spite of this, the Ewald summation method has not been extensively applied
to simulations of macromolecules in solution, where the preferred method
continues to be spherical truncation of the Coulomb potential. With the rapid
advances in computer technology and new, improved algorithms that speed the
calculation of the long-range forces without sacrificing accuracy
\cite{appel,barnes:hut,greengard,pedersen:JCP:93a,head-gordon:94},
this situation may soon
change. Thus, a discussion of the artifacts that affect the computer
simulations is timely. In this paper we focus our attention on a finite-size
effect, that is, an artifact that arises from the fact that the system being
simulated is constrained to a finite volume, $V$.  This effect can be quite
strong and, although it disappears in the limit $V\to\infty$, the convergence
to this limit is not particularly rapid.  It should be noted, however, that
volume-dependencies affect any simulation of a finite system, and are not just
an artifact of using the Ewald summation method, although the Ewald method
lends itself to a more systematic theoretical treatment.

The effect described in this paper can be seen from two different
perspectives: either as the result of the fact that constraining the system to
a finite box introduces a volume-dependent natural scale for the interaction
energy of a system of two ions; or as the dependence on the unit cell volume,
$V$, of the electrostatic interaction energy of a pair of charges immersed in
a continuum of dielectric $\varepsilon$ (with periodic boundary conditions).
These two views are expounded in sections~\ref{sec:inter-energy}
and~\ref{sec:two-ions}. In a related paper \cite{delbuono:94a} Del Buono et
al.\ demonstrate by explicit computer simulations in water
that this effect is not negligible.

The structure of the paper is as follows. In the second section we discuss the
most general form of the electric potential in a periodic box of volume $V$
and show that simple physical requirements lead to the (tinfoil) Ewald
potential augmented by an ``extrinsic'' potential, proportional to the unit
cell's dipole moment. This section does not contain new results but can shed
new light on old results that are still not quite clear in the technical
literature \cite{friedman:mezei:95a}.  In section~\ref{sec:inter-energy} the
classical electrostatics expression for the interaction energy of two point
charges in a periodic box is discussed.  It is shown that the interaction
energy coincides with the potential only if the latter satisfies the condition
that its volume average vanishes. This condition generalizes the usual ``zero
potential at infinity'' rule in infinite systems.  Section~\ref{sec:two-ions}
presents results of molecular dynamics simulations of two ions in aqueous
solution in order to illustrate the issue of volume dependence and its
importance to computer simulations. We focus on the free energy of
charging a system of two ions in a bath and show that, again, the condition of
vanishing volume average arises quite naturally.  Next we discuss a numerical
example to put into perspective the size of the effect.
Finally the relevance to computer simulations of aqueous
solutions is discussed.

\section{The lattice sums}
\label{sec:lattice-sums}

Since the seminal work of de Leeuw et al. \cite{PRSL:LPS:I,PSRL:LPS:II} it has
been known that the lattice summation method, that is, the summation of the
potentials and forces produced by an infinite number of replicas of a system
of charges (unit cell), produces not one but a whole family of potentials,
which differ in a term proportional to the dipole moment $\bbox d$ of the unit
cell and take the form \widetext
\begin{equation}
\label{eq:deleeuw}
 \Phi(\bbox x) = \sum_{\bbox n}
  {\displaystyle\mathop{\text{erfc}}\left(\alpha
        \vert\!\vert\bbox x - L\bbox n\vert\!\vert\right)
    \over
    \displaystyle \vert\!\vert\bbox x - L\bbox n\vert\!\vert}
  +
 {1\over\pi} \sum_{\bbox k \ne \bbox 0}
  {\displaystyle \exp\left(-\pi^2
  {\vert\!\vert\bbox k\vert\!\vert ^2\over\alpha^2 L^2} +
    {2\pi i\over L} \bbox{x}\mathbin{\cdot}\bbox{k} \right) \over
   \displaystyle L \vert\!\vert\bbox k\vert\!\vert ^2}
  + {\lambda\over V} \bbox x\mathbin{\cdot}\bbox d + \hbox{constant}.
\end{equation}
\narrowtext
Although for $\lambda\ne0$ this potential is not periodic due to the term
linear in $\bbox x$, it nevertheless gives rise to a periodic electric field,
which is the only constraint that can be imposed on physical grounds.
The constant of proportionality, $\lambda$, can be
related to the dielectric constant $\varepsilon_{\text{RF}}$ of a medium
surrounding a large sphere of replicas of the unit cell. In the limit of an
infinite sphere a residual dependence on this dielectric constant remains,
giving rise to this ``extrinsic'' potential
\cite{RedlackGrindlay:PCS:74}. Sometimes the choice of
$\varepsilon_{\text{RF}}$ is called the ``boundary condition,'' not to be
confused with the fact that these models impose periodic boundary conditions
on the particles.  The value $\varepsilon_{\text{RF}}= \infty $ (tinfoil or
conducting boundary) is very special since it alone results in a potential
that is periodic due to the absence of this dipolar term. All other choices
necessarily break periodicity and their use in molecular dynamics simulations
introduces some subtle questions. In what follows when we talk about the
``Ewald potential'' we mean that which results from the choice of tinfoil
boundary conditions.  If other ``boundary conditions'' are chosen another
finite-size effect related to this dipolar term appears. This will not be
discussed further in this paper.

In their excellent mathematical treatment de Leeuw et al.\
\cite{PRSL:LPS:I,PSRL:LPS:II} were concerned with the calculation of the
conditionally convergent (i.e., ill-defined) sum
\begin{equation}
\label{eq:lattice-sum}
 {1\over2} \sum_{\bbox n} \sum_{i, j}
   {q_i q_j \over
    \displaystyle \vert\!\vert\bbox x_i - \bbox x_j - L\bbox n
         \vert\!\vert}
\end{equation} that defines the potential energy of a system of charges in a
periodic box of side length $L$ (notice that for $\bbox n=\bbox0$ the diagonal
terms $i=j$ are dropped).  Formally this sum can be rewritten as
\begin{equation}
{1\over2} \sum_{i,j} q_i q_j \Phi(\bbox x_i - \bbox x_j)
\end{equation}
where
\begin{equation}
\label{eq:lattice-potential}
\Phi(\bbox x) = \sum_{\bbox n}
  {1\over \displaystyle \vert\!\vert\bbox x - L\bbox n\vert\!\vert}.
\end{equation} It is this sum, or rather a regularization of it, that we will
be concerned with.

First of all, note that if the system is electrically neutral we can add to
$\Phi$ any constant term without changing the potential energy since
\begin{equation}
\sum_{i,j} q_i q_j \times c = c \times \left(\sum_i q_i\right)^2 = 0.
\end{equation} The second observation is that $\Phi(\bbox x_i - \bbox x_j)$
diverges as $1/\vert\!\vert\bbox x_i - \bbox x_j\vert\!\vert$ when $\bbox
x_i \to \bbox x_j$. This short-distance divergence is well known and has
nothing to do with the fact that the sum in equation (\ref{eq:lattice-sum}) is
ill-defined; it is just an artifact of using point charges, which have a
divergent self-energy even in an infinite space. Thus the diagonal sum $i=j$
should be redefined to subtract this trivial divergence.  In any case, the
diagonal sum contributes only to the self-energy of the system and does not
influence the dynamics.

The third observation is the following: the sum in equation
(\ref{eq:lattice-potential}) diverges. In
their treatment de Leeuw et al.\ actually found that the divergence is
independent of $\bbox x$, and there is no effect for an electrically neutral
system. Once this divergence is removed what remains is conditionally
convergent.  We should mention that there seems to be a confusion about this
point: the \textsl{only} requirement that needs to be made to obtain sensible
results is that the system be electrically neutral and not, as has been
mentioned in the literature \cite{friedman:mezei:95a}, that the net dipole
moment vanish as well. If there is a net dipole moment the sum in equation
(\ref{eq:lattice-potential}) is conditionally convergent (not divergent),
which means that there is not a unique answer. This nonuniqueness reflects the
fact that some
properties of a system depend not only on the intrinsic characteristics of the
system but also on the shape and other boundary conditions.
Due to its long-range nature, the electrostatic energy is a prime
example of this dependence. A clear discussion of this point, albeit quite
technical, can be found in \cite{RedlackGrindlay:PCS:74}.

Since
$\Phi(\bbox x - \bbox u)$ is the electric potential produced by a unit charge
at $\bbox u$ (plus all its infinite copies), we might start by requiring that
$\Phi$ satisfies Poisson's equation \cite{jackson}
\begin{eqnarray}
\label{eq:poisson-r3}
\nabla^2 \Phi(\bbox x - \bbox u) = -4\pi \delta_p(\bbox x - \bbox u)
\end{eqnarray}
and that $\Phi$ be a periodic function: $\Phi(\bbox x + L \bbox
n) = \Phi(\bbox x)$ (here $\delta_p$ is the periodic Dirac $\delta$-function).
This is, however, not consistent, as integrating both sides over the unit cell
shows. The left hand side gives identically zero, as can be seen by using
Stoke's theorem to convert the volume integral to a surface integral and the
fact that the electric field is periodic, while the right hand side
does not vanish. This is simply a consequence of the well known fact that in a
periodic system the net charge must vanish. We can easily satisfy the
neutrality requirement by adding a homogeneous background charge density that
exactly cancels the unit charge in equation (\ref{eq:poisson-r3}):
\begin{eqnarray}
\label{eq:poisson}
\nabla^2 \Phi(\bbox x-\bbox u) &=& -4\pi  \left(\delta_p(\bbox x - \bbox u) -
  {1\over V}\right) \nonumber \\
  &=& -4\pi  \sum_{\bbox k \ne \bbox
  0} e^{{2\pi i\over L} (\bbox x - \bbox u)\mathbin{\cdot}\bbox k}
\end{eqnarray}
where we have also written the right hand side in Fourier space (here
$V=L^3$).

With hindsight we can find a particular solution of equation
(\ref{eq:poisson}): Ewald's potential with conducting, or tinfoil, boundary
conditions, \cite{ewald:21,allen:tildesley,PRSL:LPS:I,PSRL:LPS:II}, which
corresponds to $\lambda=0$ in equation~(\ref{eq:deleeuw}): \widetext
\begin{equation}
\label{eq:pewald}
{\Phi'_{_{\text{Ew}}}}(\bbox x) =
 \sum_{\bbox n}
   {\displaystyle\mathop{\text{erfc}}\left(\alpha
        \vert\!\vert\bbox x - L\bbox n\vert\!\vert\right)
    \over
    \displaystyle \vert\!\vert\bbox x - L\bbox n\vert\!\vert}
  +
 {1\over\pi} \sum_{\bbox k \ne \bbox 0}
  {\displaystyle \exp\left(-\pi^2 {\vert\!\vert\bbox k\vert\!\vert
      ^2\over\alpha^2 L^2} + {2\pi i\over L} \bbox x\mathbin{\cdot}\bbox k
  \right) \over \displaystyle L \vert\!\vert\bbox k\vert\!\vert ^2}.
\end{equation}
\narrowtext As written this potential has a flaw: it depends on
the arbitrary parameter $\alpha$. This can be seen most easily by considering
the mean value of ${\Phi'_{_{\text{Ew}}}}$ over the unit cell $V$:
\begin{eqnarray}
{1\over V} \int\limits_V d^3 x\, {\Phi'_{_{\text{Ew}}}}(\bbox x) & = &
  \sum_{\bbox n}
 {1\over V}\int\limits_V d^3 x\,
   {\displaystyle\mathop{\text{erfc}}\left(\alpha
      \vert\!\vert \bbox x - L\bbox n \vert\!\vert\right)
    \over
    \displaystyle \vert\!\vert \bbox x - L\bbox n
          \vert\!\vert} \nonumber \\ &=&
 {1\over V}\int d^3 x\,
    {\displaystyle\mathop{\text{erfc}}(\alpha\vert\!\vert \bbox
      x \vert\!\vert)
      \over\displaystyle \vert\!\vert \bbox x \vert\!\vert} =
  {1\over L} {\pi\over\displaystyle (\alpha L)^2}. \nonumber
\end{eqnarray}
Subtracting this mean value results in an $\alpha$-independent potential,
\begin{equation}
\label{eq:ewald}
{\Phi_{_{\text{Ew}}}}(\bbox x) = {\Phi'_{_{\text{Ew}}}}(\bbox x) -
   {1\over L} {\pi\over\displaystyle (\alpha L)^2}
\end{equation}
which, furthermore, has the property that
\begin{equation}
{1\over V} \int\limits_V d^3 x\, {\Phi_{_{\text{Ew}}}}(\bbox x) = 0.
\end{equation}
Applying the laplacian $\nabla^2$ to ${\Phi_{_{\text{Ew}}}}$ one verifies,
after a few manipulations, that it is a solution of equation
(\ref{eq:poisson}).

Thus we can write our sought-after potential $\Phi$ in the form
\begin{equation}
\Phi(\bbox x) = {\Phi_{_{\text{Ew}}}}(\bbox x) + \phi'(\bbox x)
\end{equation}
where now $\phi'$ satisfies the homogeneous Laplace's equation
\begin{equation}
\label{eq:laplace}
\nabla^2 \phi'(\bbox x) = 0.
\end{equation}
In Appendix \ref{appx:homogeneous} the most general solution of the
homogeneous problem is derived. The general form
found by de Leeuw et al., equation~(\ref{eq:deleeuw}), is recovered.
In the remainder of the paper we choose the solution to the homogeneous
problem equal to zero, which corresponds to setting $\lambda=0$ in
equation~(\ref{eq:deleeuw}).

\section{Interaction energy of two ions. Continuum electrostatics approach}
\label{sec:inter-energy}

If two ions are placed in a periodic box filled with a solvent and the
separation between them is large compared with typical molecular correlation
lengths,  the free energy of the system is given by the classical continuum
electrostatic energy. The effect of the solvent is merely to renormalize the
vacuum interaction energy (dielectric screening), and the complicated many
body problem is reduced to a simpler two body problem.
In this section we focus on this interaction energy from the point of view of
continuum electrostatics, and in the next section we present a microscopic
treatment of the same quantity. For large enough separations, of course, both
treatments should give the same results.

Let's therefore consider a system of two point
charges, together with their respective neutralizing backgrounds, in a cubic
box with periodic boundary conditions filled with a dielectric medium of
dielectric constant $\varepsilon$.
It will be shown that continuum electrostatics predicts
that the interaction energy
shows a rather strong finite-size effect, i.e., a dependence on the linear
dimension $L$ of the box. In what follows we set $\lambda=0$ so that the
electric field is derived from a well-defined periodic potential.

In general we can expand $\bbox E$ as a Fourier series
\begin{equation}
\label{eq:fourier}
\bbox E(\bbox x) = \sum_{\bbox k\ne\bbox0}
 \bbox E_{\bbox k} \, e^{ {2\pi i \over L} \bbox k\mathbin{\cdot}\bbox x }
\end{equation}
excluding, as discussed above, the $\bbox k=\bbox0$ term.
The other terms can all be obtained from equation (\ref{eq:poisson}):
\begin{equation}
 -4\pi^2 {\displaystyle\vert\!\vert \bbox k \vert\!\vert ^2
  \over\displaystyle L^2} \phi_{\bbox k} = -4\pi \rho_{\bbox k}
  \Rightarrow
  \phi_{\bbox k} = {\displaystyle L^2 \rho_{\bbox k}\over\displaystyle
   \pi\vert\!\vert \bbox k \vert\!\vert ^2},
\end{equation}
\noindent together with the Fourier space version of $\bbox E = -\nabla\phi$:
\begin{equation}
 \bbox E_{\bbox k} = -{2\pi i\over L} \bbox k \,\phi_{\bbox k}.
\end{equation}
The term $\phi_{\bbox 0}$ is not fixed by equation (\ref{eq:poisson}); since it
doesn't contribute to the physical observables (like $\bbox E$) this term is
completely arbitrary. In particular, the Ewald potential
${\Phi_{_{\text{Ew}}}}$ corresponds to $\phi_{\bbox0}=0$.

Consider now a system consisting of two charge distributions, $\rho_1$ and
$\rho_2$, which are well separated, i.e., $\rho_j$ is zero outside
of a small neighborhood of point $\bbox u_j$, and the two neighborhoods do not
overlap. The energy of the system can be written (assuming a medium with
dielectric constant $\varepsilon$)
\begin{equation}
E = {\varepsilon\over 8\pi}\int\limits_{V} d^3x\,
   \vert\!\vert \nabla\phi \vert\!\vert^2
\end{equation}
where $\phi$ satisfies the equation
\begin{equation}
\label{eq:two-rhos}
\nabla^2\phi(\bbox x) = -{4\pi\over\varepsilon}
 \left( \rho_1(\bbox x) + \rho_2(\bbox x)\right).
\end{equation}
For simplicity we assume that
\begin{equation}
Q_j = \int\limits_{V} d^3x\, \rho_j \ne 0, \qquad
Q_1 + Q_2 = 0
\end{equation}
otherwise a further neutralizing background must be
added to~(\ref{eq:two-rhos}).
We want to extract from the total energy $E$ the interaction piece. In an
infinite system ($L=\infty$) this is done by subtracting from $E$ the terms
\begin{equation}
E_j = {\varepsilon\over 8\pi}\int\limits_{V} d^3x\,
  \vert\!\vert \nabla\phi_j \vert\!\vert^2
\end{equation}
where
\begin{equation}
\nabla^2\phi_j(\bbox x) = -{4\pi\over\varepsilon}\rho_j(\bbox x).
\end{equation}
In a periodic box, as discussed previously, this equation must be replaced by
\begin{equation}
\label{eq:poisson-again}
\nabla^2\phi_j(\bbox x) = -{4\pi\over\varepsilon}\left(\rho_j(\bbox x) -
 {Q_j\over V}\right).
\end{equation}
Subtracting $E_1+E_2$ from $E$ we obtain the interaction energy,
$E_{\text{int}}$
\begin{equation}
E_{\text{int}} = {\varepsilon\over4\pi}\int\limits_{V} d^3x\,
\nabla\phi_1\mathbin{\cdot}\nabla\phi_2.
\end{equation}
Integrating by parts and using equation (\ref{eq:poisson-again})
we can rewrite this as
\begin{eqnarray}
E_{\text{int}}
 &=& \int\limits_{V} d^3x\, \left(\rho_1(\bbox x) - {Q_1\over V}\right)
     \phi_2(\bbox x) \nonumber \\
 &=& \int\limits_{V} d^3x\, \rho_1(\bbox x) \left( \phi_2(\bbox x) -
     {1\over V} \int\limits_{V} d^3y\, \phi_2(\bbox y)\right).
\end{eqnarray}
Notice that
\begin{equation}
{1\over V}\int\limits_{V} d^3y\, \phi_2(\bbox y)
\end{equation}
is nothing else than the $\bbox k=\bbox0$ term in the Fourier decomposition,
sometimes also called the \textsl{zero mode}.
This simple argument shows that to calculate the electrostatic interaction
energy between two charge distributions the $\bbox k=\bbox 0$
Fourier mode of the
potential must be removed.

Specializing to the case of two point charges, of charge $+q$ and
$-q$, situated at $\bbox u_1$ and $\bbox u_2$, respectively, we find
\begin{equation}
E_{\text{int}} = - {\displaystyle q^2\over\varepsilon}\,
   {\Phi_{_{\text{Ew}}}}(\bbox u_1 - \bbox u_2)
\end{equation}
where ${\Phi_{_{\text{Ew}}}}$ is defined by equation (\ref{eq:ewald}) so that
its integral over the unit cell vanishes identically.

\section{Interaction energy of two ions. Free energy approach}
\label{sec:two-ions}

In this section we show again, but from a statistical mechanics perspective,
that the interaction energy of two ions in solution coincides with the
potential if and only if the latter satisfies that its volume average
vanishes.

Consider two ions of charges $q_1$ and $q_2$ immersed in a molecular solvent
inside a box of volume $V$. Under very general conditions it is possible to
write the free energy of solvation of the ions as a quadratic function of the
charges (see Appendix \ref{sec:free-energies} and reference \cite{gaussian}):
\begin{eqnarray}
\label{eq:f-one}
 F_{\text{solv}}(q_1,q_2) &=& F_{\text{solv}}(q_1) +
 F_{\text{solv}}(q_2) + q_1 q_2 \Phi(\bbox x_1 - \bbox x_2)
   \nonumber \\
 & & \quad {} +
                 q_1 q_2 W(\bbox x_1 - \bbox x_2)
\end{eqnarray}
where $F_{\text{solv}}(q)$ is the free energy of solvation of one ion
immersed in the solvent when the charge in the other is set to zero (there is
still a dependence on the cavity produced by the other ion, but it is
negligible at large separations of the charges).
Here $\bbox x_i$ ($i=1$ and 2) are the positions of the ions inside
the box and $\Phi$ is the the interaction energy in the absence of solvent.
The quantity $W(\bbox x_1 - \bbox x_2)$ is the solvent contribution to the
effective interaction
energy and can be written as (Appendix \ref{sec:free-energies})
\widetext
\begin{equation}
\label{eq:w-one}
W(\bbox x_1 - \bbox x_2) = -{1\over\beta}
 {\partial^2\ln Z\over\partial q_1 \partial q_2} =
 -\beta \biggl(\langle {\phi_{\text{solv}}}(\bbox x_1)
{\phi_{\text{solv}}}(\bbox x_2)\rangle -
  \langle{\phi_{\text{solv}}}(\bbox x_1)\rangle
  \langle{\phi_{\text{solv}}}(\bbox x_2)\rangle \biggr)
\end{equation}
\narrowtext
\noindent where ${\phi_{\text{solv}}}(\bbox x)$ is the total electrostatic
potential produced by the solvent atoms at the position $\bbox x$. The mean
values are computed with the ionic charges set to zero.  As written, $W$
depends on the size of the ionic cavities; however, this dependence is
expected to be of short range. If the separation $\vert\!\vert \bbox x_1 -
\bbox x_2 \vert\!\vert$ between the ions is larger than the corresponding
short-range correlation length, it should be a good approximation to replace
the mean values in equation (\ref{eq:w-one}) by mean values computed for zero
charge and \textsl{in the absence of any cavity}, which we will denote by the
subscript `nc' \widetext
\begin{equation}
 W(\bbox x_1 - \bbox x_2) \simeq
 -\beta \biggl(\langle {\phi_{\text{solv}}}(\bbox
 x_1){\phi_{\text{solv}}}(\bbox x_2)\rangle_{\text{nc}} -
  \langle{\phi_{\text{solv}}}(\bbox x_1)\rangle_{\text{nc}}
  \langle{\phi_{\text{solv}}}(\bbox x_2)\rangle_{\text{nc}}\biggr).
\end{equation}
\narrowtext
By translation invariance the averages $\langle{\phi_{\text{solv}}}(\bbox
x)\rangle_{\text{nc}}$ are position independent and, moreover, vanish in most
cases. We will keep them, however, for the sake of generality.
If the simulation volume $V$ were infinite we would
expect this correlation function to decay to zero as $\vert\!\vert \bbox
x_1-\bbox x_2 \vert\!\vert$ approaches infinity.  From continuum
electrostatics \cite{jackson}, we expect the decay to be given by the
asymptotic formula
\begin{equation}
\label{eq:asymp-infinite-vol}
W(\bbox x_1 - \bbox x_2) \simeq \left({1\over\varepsilon}-1\right)
  {1\over\displaystyle\vert\!\vert \bbox x_1 - \bbox x_2 \vert\!\vert}
\end{equation}
where $\varepsilon$ is the solvent dielectric constant. Note that the physical
decay of the correlations selects a ``natural'' scale for the interaction
energy: it should be zero at infinite separation.
In a finite, periodic box, continuum electrostatics leads us to expect that the
Coulomb potential in equation~(\ref{eq:asymp-infinite-vol}) should be
replaced by its appropriate periodic version: the Ewald potential
\begin{equation}
\label{eq:asymp-finite-vol}
W(\bbox x_1 - \bbox x_2) \simeq \left({1\over\varepsilon}-1\right)
  {\Phi_{_{\text{Ew}}}}(\bbox x_1 - \bbox x_2).
\end{equation}
In the infinite volume case we require that the potential go to zero at
infinite separation. Since the ions in a finite box cannot be separated to an
infinite distance, what is the equivalent ``natural'' scale for the
interaction energy? A simple argument provides an answer, which is borne out
by a further analysis. The interaction energy is that part of
the energy that depends on the positions of both ions;
therefore, any part of the potential in (\ref{eq:asymp-finite-vol}) that does
not depend on the positions of the ions must be removed.  One way to do this
is to set the $\bbox k=\bbox 0$ Fourier component of the potential to
zero. All the other modes depend in an essential way on the positions of the
ions and are left untouched.  Thus we are led to the conclusion that the
``natural'' choice for periodic boundary conditions is that which has no
$\bbox k=\bbox0$ Fourier mode. A more rigorous way to arrive at this same
conclusion is to observe that the integral over the simulation box vanishes
(see Appendix~\ref{sec:free-energies}):
\widetext
\begin{equation}
{1\over V}\int\limits_V d^3u\,
\biggl(\langle {\phi_{\text{solv}}}(\bbox u){\phi_{\text{solv}}}(\bbox
  0)\rangle_{\text{nc}} -
  \langle{\phi_{\text{solv}}}(\bbox u)\rangle_{\text{nc}}
  \langle{\phi_{\text{solv}}}(\bbox 0)\rangle_{\text{nc}} \biggr) = 0.
\end{equation}
\narrowtext
This integral is in fact the $\bbox k=\bbox0$ Fourier mode of this
two-point correlation function.

Since the Fourier modes depend on the simulation volume, the requirement that
the $\bbox k=\bbox0$ mode be zero introduces a subtle volume dependence.
The next section shows with an example that this dependence can be large.

\section{Interaction energy of two ions. Explicit simulation}
\label{sec:numer}

In this section we present the results of molecular dynamics simulations of
two ions in aqueous solution using the Ewald summation method.  After
evaluating the charging free energy the volume dependence is discussed.

In a large dielectric medium the contribution of the solvent to the free
energy of charging, equation~(\ref{eq:asymp-finite-vol}), almost exactly
cancels the direct interaction energy (i.e., interaction in the absence of
solvent).  In order to correctly model the dielectric response of the solvent
in molecular simulations this balance needs to be accurately evaluated.

{}From equation~(\ref{eq:f-one}), the free energy of solvation of two charged
particles can be related to several terms that are easily identifiable with
particular interactions within the system.  The first two terms represent the
free energy of solvation of each ion in solution when the other ion has charge
zero; the second term is the direct interaction contribution, and the last
term is related to the solvent dielectric shielding.  This last term is
what will be obtained from simulations.  There are at least two ways in which
the solvent dielectric shielding contribution can be explicitly evaluated: one
is in the absence of the charges through a solvent correlation function (see
equation~(\ref{eq:w-one})); the other is in presence of the charges by
relating the dielectric shielding to the average
electrostatic potentials of the systems with two and one charged particles
(see Appendix \ref{sec:free-energies}).  The simulations discussed in this
section were designed to use the second approach.

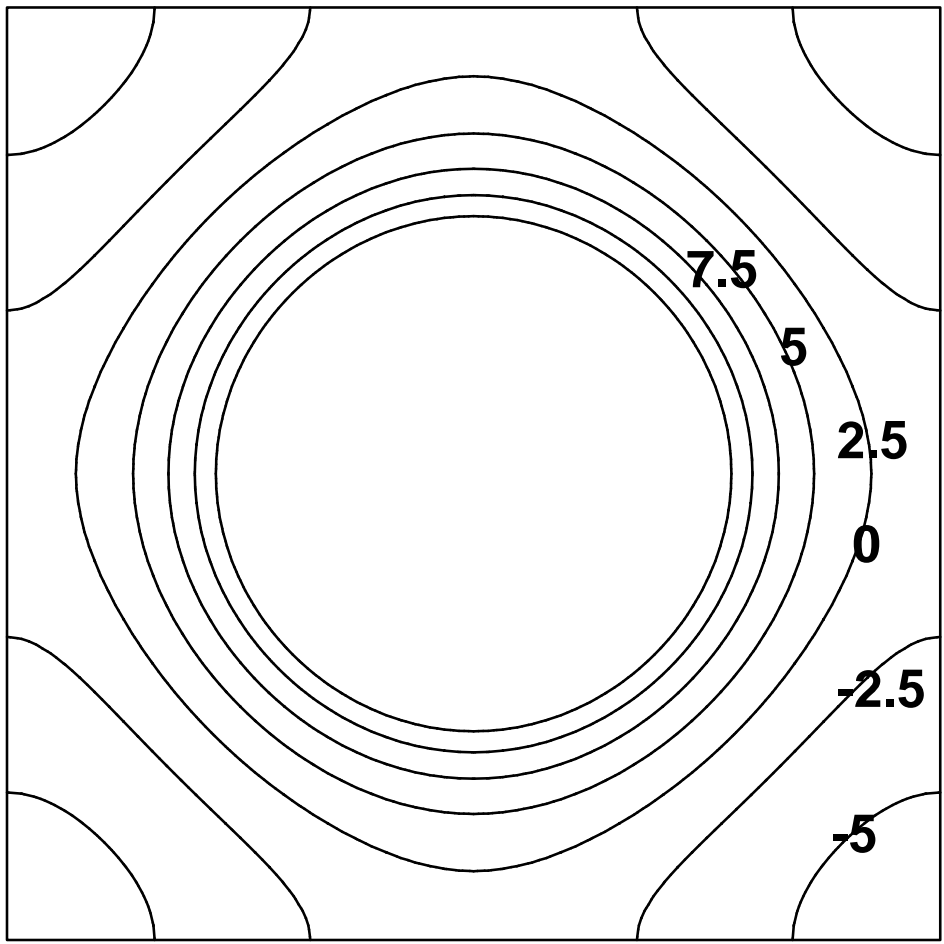
\begin{figure}
\caption{\label{fig:ew-contour}%
Contour plot of the Ewald potential. The labels correspond to the levels in
kcal/mol$\cdot$au, for a simulation box of 32~\AA.
The contours lie all in the plane $z=0$.}
\end{figure}

Two oppositely charged chloride-like ions were situated on an axis
perpendicular to two faces of the unit cell 10~\AA~apart in a 32~\AA~cubic box
with 1103 SPC water molecules (the orientation is important as the Ewald
potential is not isotropic, see Fig.~\ref{fig:ew-contour}).
Three 200 picosecond molecular dynamics
simulations were performed at room temperature for systems with the two
charged particles, and systems with only one of the particles charged. The
average electrostatic potentials due to the solvent were evaluated at the
cavity sites and the free energy of charging was calculated from the
difference of the average potentials using the formula
\begin{equation}
\Delta F_{0\to q, 0\to-q} = \frac{1}{2} \left(
  \langle {\phi_{\text{solv}}}(\bbox x^{+})\rangle_{q,-q} -
  \langle {\phi_{\text{solv}}}(\bbox x^{-})\rangle_{q,-q} \right)
\end{equation}
where $\bbox x^{+}$ and $\bbox x^{-}$ are, respectively, the positions of the
positive and negative ions (see Appendix \ref{sec:free-energies}).
More details about the methodology and
additional studies of the effect of the truncation scheme on the dielectric
shielding can be found in a related paper~\cite{delbuono:94a}.

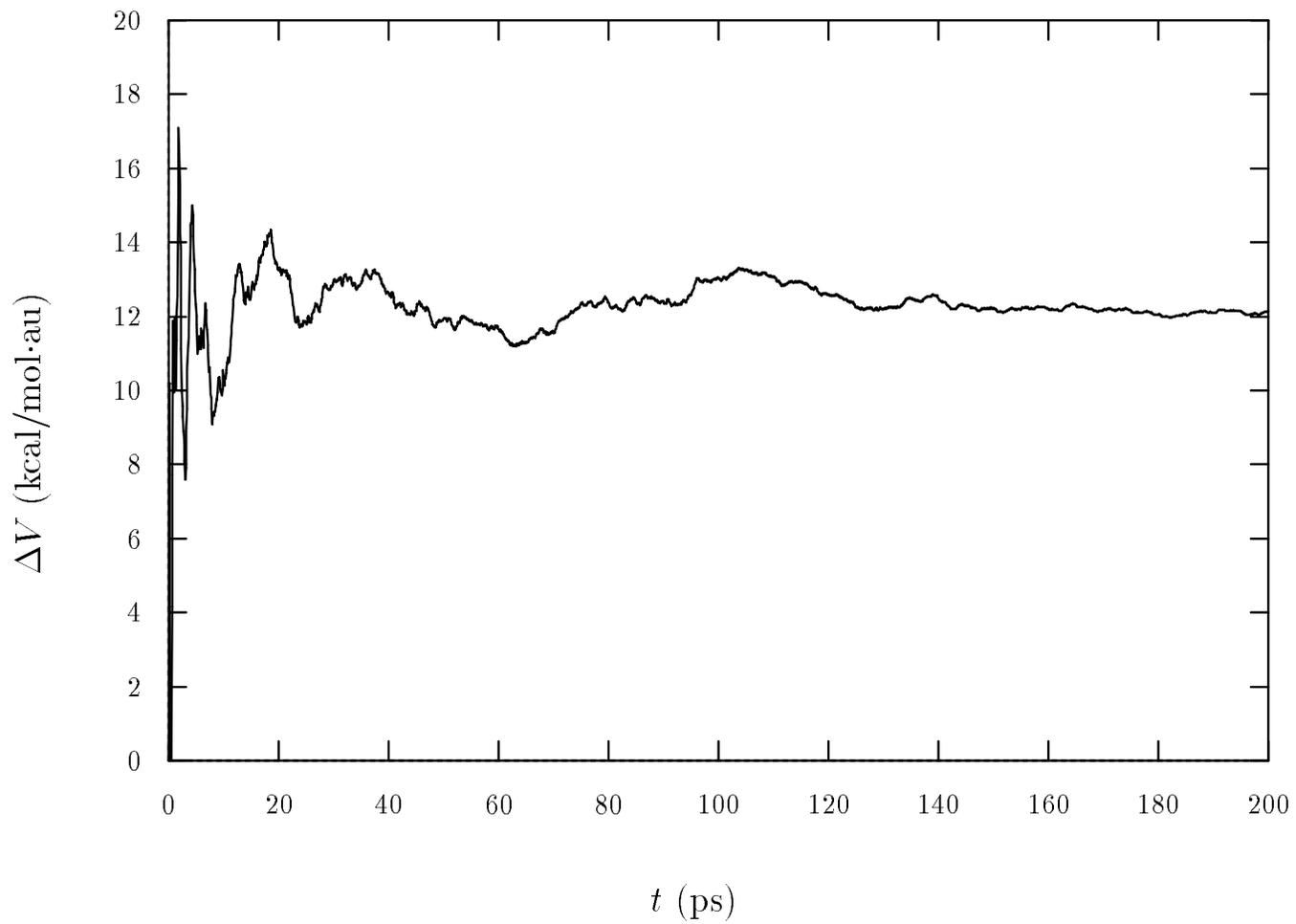
\begin{figure}
\caption{\label{fig:running}%
Running average of the difference in the electrostatic potential produced by
the solvent in the charged and uncharged states.}
\end{figure}

Figure~\ref{fig:running}
shows the running average of the difference between the
electrostatic potentials due to the solvent at the ionic positions.
In order to obtain the free
energy of charging due to the
solvent
the number obtained is divided by two, yielding $-6.1$~kcal/mol.  The direct
interaction energy of the ions in the absence of the solvent is
$-6.12$~kcal/mol; therefore the near cancellation of the direct interaction by
the
interaction with the solvent is observed, as is expected for a high
dielectric solvent (water).

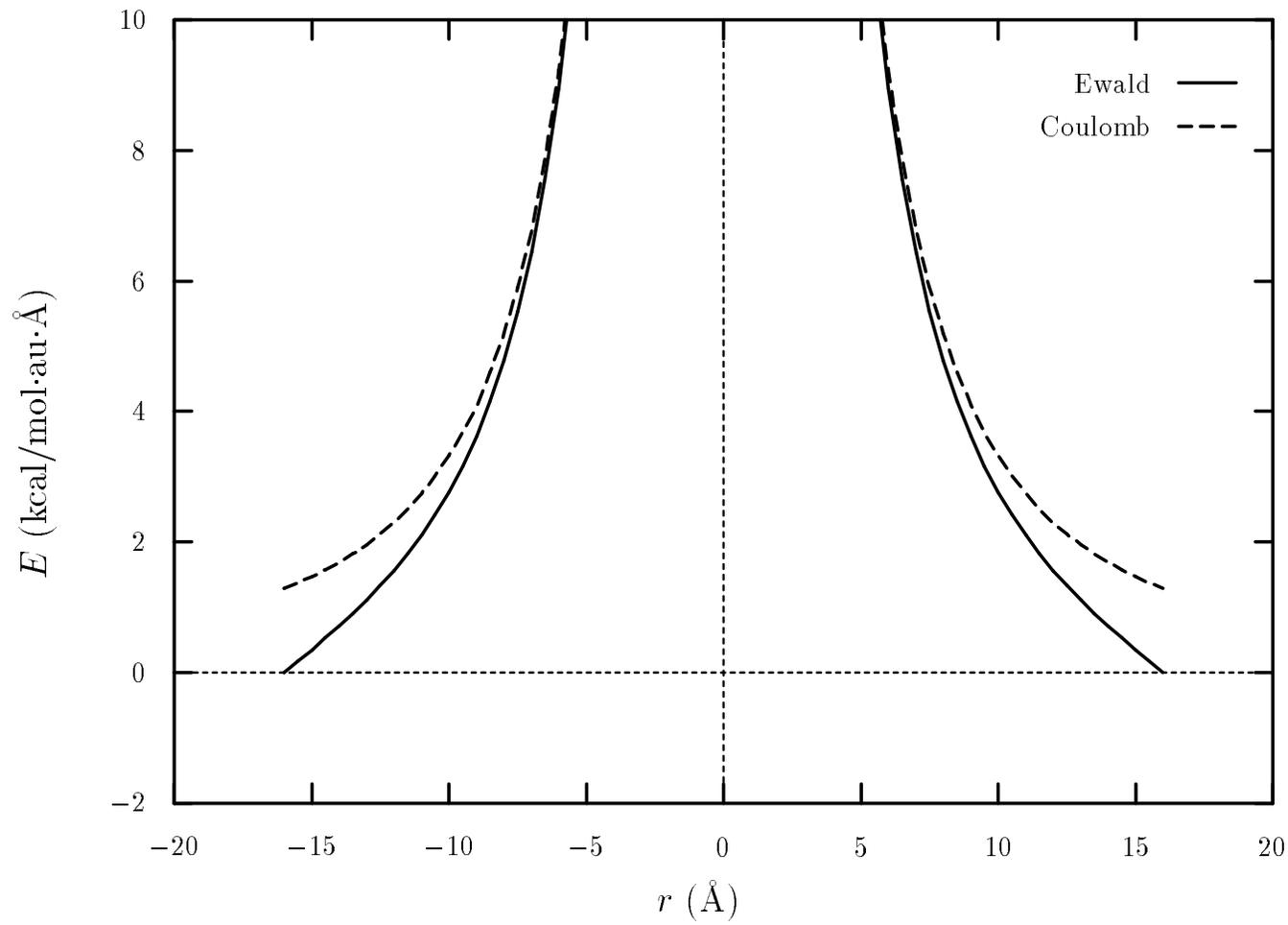
\begin{figure}
\caption{\label{fig:ew-cou-for}%
Comparison between Ewald (full curve) and Coulomb (dashed
curve) fields for a unit charge at the origin in a box of side length
32~\AA.}
\end{figure}

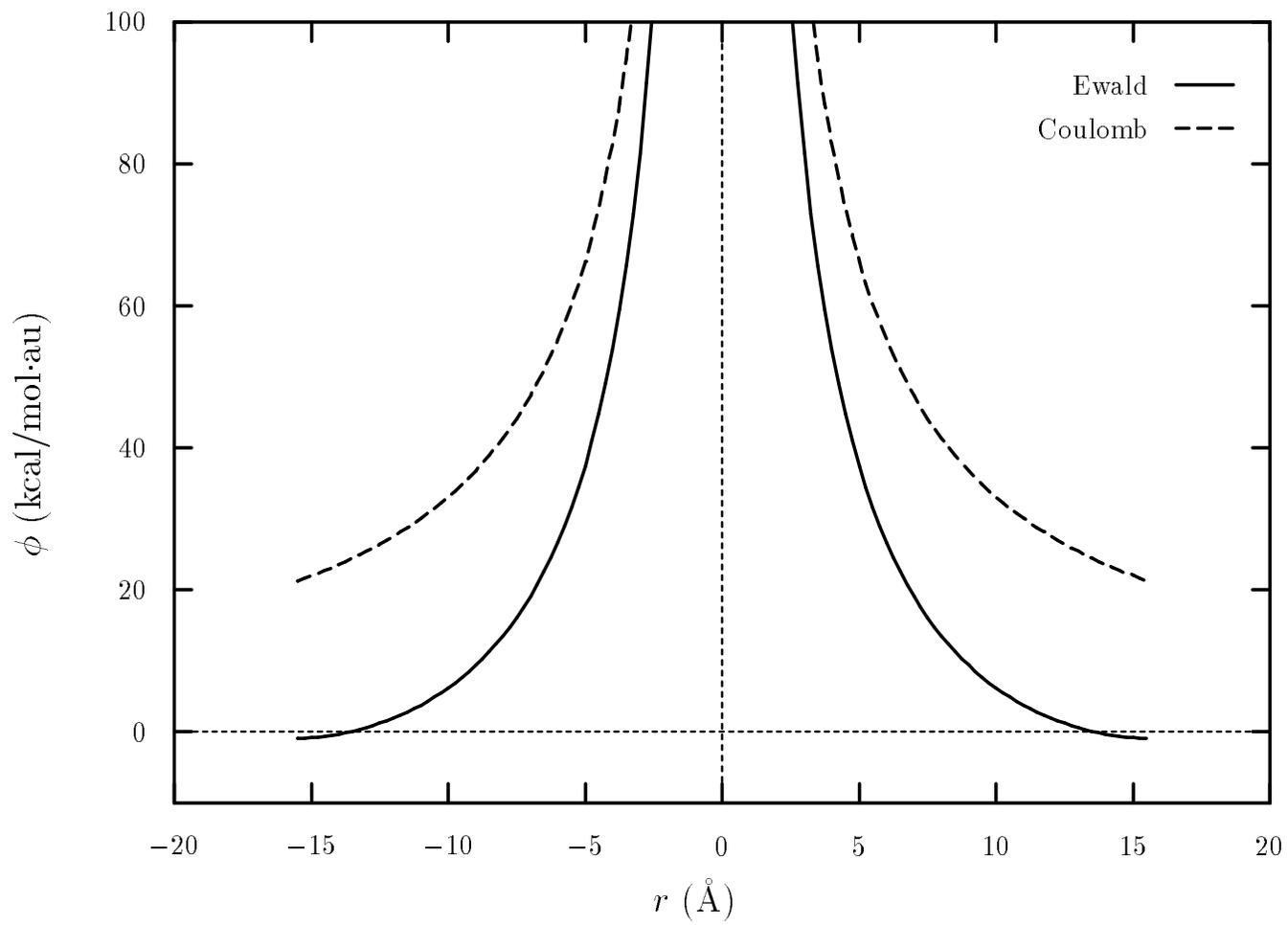
\begin{figure}
\caption{\label{fig:ew-cou-pot}%
Comparison between Ewald (full curve) and Coulomb (dashed
curve) potentials for a unit charge at the origin in a box of side length
32~\AA.}
\end{figure}

It is relevant at this point to make a comparison between the finite volume
direct interaction energy in the absence of solvent,
$-6.12$~kcal/mol, and the
infinite volume direct interaction energy, $-32.77$~kcal/mol.
What produces this large difference? If we compare the fields produced by the
Coulomb and Ewald potentials (Fig.~\ref{fig:ew-cou-for}), a much smaller
difference is found. On the other hand, the numbers just quoted refer to the
(unmodified) Coulomb and the particular Ewald potential that satisfies the
condition of vanishing volume average (no zero mode). These are compared in
Fig.~\ref{fig:ew-cou-pot}. The Ewald potential is shifted down by a
considerable amount with respect to the Coulomb potential. Besides this shift
there are other differences that appear because of the requirement of
periodicity (note that the Coulomb potential does not meet the boundaries
smoothly).  These distortions are also of interest and will be the
subject of a future publication.

We can estimate the importance of this shift downward with respect
to the Coulomb potential as a function of the simulation volume as
follows. Neglecting, to a first approximation, the distortions mentioned in
the previous paragraph, the amount of this shift should be given by the zero
mode, or volume average, of the Coulomb potential. Actually, since the Coulomb
potential is not periodic this computation is ill-defined; however, we can
replace it by its simplest periodic extension, the ``minimum image''
potential. Its volume average is easy to calculate and we obtain
\begin{eqnarray}
{1\over V} \int\limits_V d^3 x \, {1\over\vert\!\vert \bbox x \vert\!\vert}
  &=& {2\over L} \int\limits_0^1 dx \int\limits_0^1 dy \int\limits_0^1 dz
   \, {1\over\displaystyle\sqrt{x^2 + y^2 + z^2}} \nonumber \\
 &\simeq&  {2\over L} \times 1.19
\end{eqnarray}
where $L$ is the linear dimension of the simulation box ($V=L^3$).
For the case considered above ($L=32$~\AA) this gives about 24~kcal/mol, a
substantial portion of the difference ($32.77 - 6.12 = 26.65$) between the
infinite volume limit and the correct finite volume answer.
This justifies, a posteriori, the neglect of the shape distortions as a first
approximation. Thus, to a large extent, we can
say that the Ewald potential, equation (\ref{eq:ewald}),
is ``just'' the Coulomb potential but shifted
down to guarantee that its mean value over the finite box is zero.
This shift introduces a volume dependence that, as we have seen, is not
negligible.

\section{Discussion}
\label{sec:discussion}

In the previous sections we have seen how a strong finite-size effect arises,
which shifts the interaction energy of two point
charges by a nonnegligible amount for the size of boxes that are customary at
present. We have also discussed why it is mostly due to a shift downwards of
the Ewald with respect to the Coulomb potential.

As shown by Del Buono et al.\ \cite{delbuono:94a}, when this effect is
taken into account different calculations of the dielectric response give
results that are consistent with each other and with experimental
observations. This is expected for intensive quantities, like the dielectric
constant, which are not strongly dependent on the shape of the sample
\cite{AdelmanDeutch:ACP:xxxi,Stell:ACP:xlviii}.
For the calculation of extensive quantities, however, the finite size effects
described in this paper can produce results that are not directly comparable
with experiment. A particularly interesting example is provided by the
calculation of pK$_a$ values
\cite{delbuono:94,bashford:karplus,Warshel:86,Honig:JMB:93,Gilson:JMP:94}, or
rather, the
shift, of titratable residues in biological macromolecules. Since these
shifts can be related to the interaction energy of the set of titratable
residues \cite{bashford:karplus},
we expect a dependence on the size of the simulation box.

The constant field in equation (\ref{eq:deleeuw}) introduces problems for
the consistent simulation of the system unless $\lambda=0$.
The problems arise
because of the fact that the system ``remembers'' the unit cell from which it
was constructed by infinite replication, and, as a consequence, this field
does not depend continuously on the position of the charges: there is a finite
jump whenever a charge crosses the boundaries of the unit cell.
It is interesting to notice that if, instead of point charges point dipoles
are used, these discontinuities do not arise, since the total dipole moment is
independent of the position of the dipoles \cite{KMS:JCP:94}.
However, in simulations using
explicit atom models these jumps are bound to occur and it is not clear what
the correct implementation of the dynamics is in that case. Until this point
is clarified it seems best to set $\lambda = 0$ (see, however, reference
\cite{Caillol:JCP:94} for a possible solution).

The appearance of strong finite-size effects is not surprising, given the
long-range nature of the electrostatic forces.
Many earlier simulations were focused either on radial distribution functions,
$g(r)$, which show remarkable insensitivity to the detailed treatment of the
electrostatic forces, or on the computation of the dielectric constant, which
is an intrinsic measure of the response of the solvent and not
affected by these finite-size effects. It is well known, however, that the
Kirkwood $G_{\text{K}}$-factor, and its detailed relation to the dielectric
constant of the medium, is strongly dependent on the choice of boundary
conditions \cite{neumann:dielectric-sums}.
Nowadays many
simulations, particularly those concerned with biological molecules, attempt
to compute (differential) free energies, and it is reasonable to expect that
these calculations will be dependent on the size of the simulation box through
the volume dependence of the electrostatic potential. How strong this
dependence really is in a realistic simulation is a question that we will
attempt to answer in the future.

After this work was completed we received a preprint by Hummer, Pratt and
Garc\'\i a \cite{hummer:95a}
where a related finite-size effect was reported that affects the
calculation of free energies of hydration of single ions.

\section*{Acknowledgments}
We thank particularly David A. Coker for critical
reading of the manuscript. We also thank J\"urgen Schnitker for kindly sending
us his preprints on simulations with the Ewald potential. This work has been
supported in part by grants from the National Institute of Health (GM30580)
and the Columbia University Center for Biomolecular Simulations (NIH P41
RR06892).

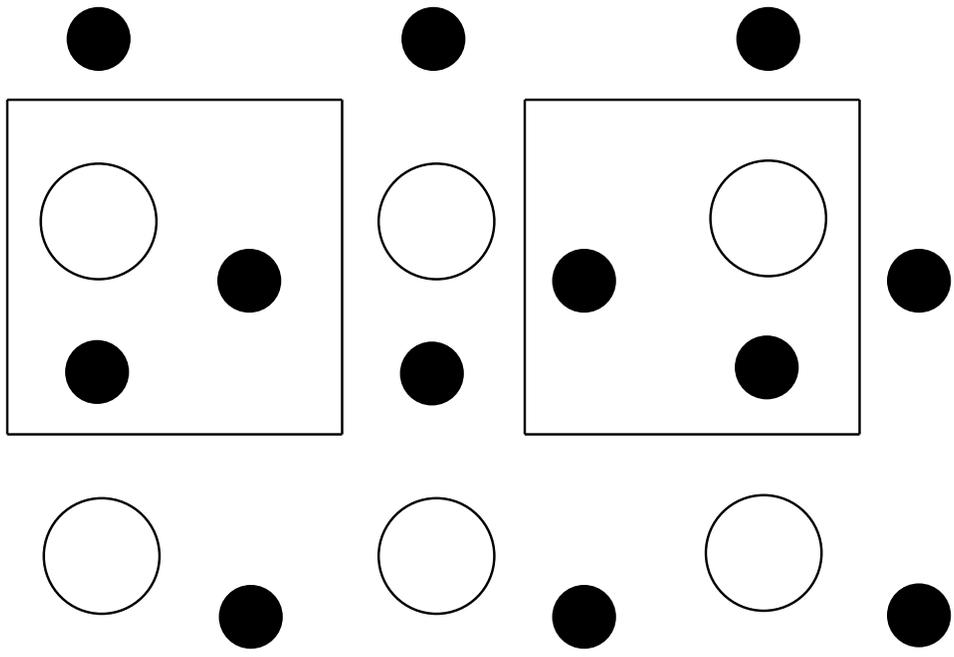
\begin{figure}
\caption{\label{fig:balls}%
Two choices of unit cell in a periodic system.}
\end{figure}

\appendix
\iftrue
\section{}
\label{appx:homogeneous}

In this appendix we derive the most general form of a solution to the
homogeneous equations
\begin{equation}
\label{eq:div-free}
\nabla\cdot \bbox E = 0 \hbox{~~and~~} \nabla\times\bbox E = 0
\end{equation}
for the electrostatic field in a periodic box of linear dimension $L$.

Algebraic topology \cite{Geroch,BottTu} can be used to prove that the only
\textit{periodic} solutions to (\ref{eq:div-free}) are constant fields. Note
that if
\begin{equation}
-\nabla \phi'(\bbox x) = \bbox E_0,
\end{equation}
for a constant $\bbox E_0$, then $\phi'$ must have the form
\begin{equation}
\phi'(\bbox x) = - \bbox x\mathbin{\cdot}\bbox E_0 + \hbox{$\bbox
  x$-independent terms}
\end{equation}
which is not a periodic potential.

The explicit form of $\bbox E_0$ given by de Leeuw et al.\
\cite{PRSL:LPS:I} can be recovered by the following argument.
First, we make the natural assumption that $\bbox E_0$ depends only
on the distribution of charge inside the unit cell, that is, we assume there
are no charges ``at infinity.'' Moreover, we assume this dependence is linear
and that the ``charges inside the unit cell'' include the homogeneous
background (if present). Introducing the Green's function $\bbox G(\bbox
x,\bbox u)$ we can then write
\begin{equation}
\label{eq:ezero}
\bbox E_0 = \int\limits_V d^3 u\, \bbox G(\bbox x, \bbox u) \,
\left( \rho(\bbox u) - {1\over V}
  \int\limits_V d^3 v \, \rho(\bbox v) \right).
\end{equation}
Below we will see how the form of $\bbox G(\bbox x,\bbox u)$ is constrained by
the requirement that $\bbox E_0$ be constant; in the end we will recover the
expression obtained by de Leeuw et al.

Notice that since
\begin{equation}
\int\limits_V d^3 u\, \left(\rho(\bbox u) - {1\over V}
  \int\limits_V d^3 v \, \rho(\bbox v) \right) = 0
\end{equation}
for any charge distribution $\rho$, we are free to add to $\bbox G(\bbox x,
\bbox u)$ an arbitrary $\bbox u$-independent term.
Furthermore, because of translational invariance, $\bbox G$ can only depend on
relative coordinates:
$\bbox G(\bbox x, \bbox u) = \bbox G(\bbox x -\bbox u)$.
Taking derivatives with respect to $x^j$ on both sides of
equation (\ref{eq:ezero}), one can prove that $\bbox G$ must be of the form
\begin{equation}
\bbox G(\bbox x, \bbox u) = {\lambda\over V} (\bbox x - \bbox u) + \bbox G_0 .
\end{equation}
Here we have again assumed that the symmetries of the cubic unit cell are
preserved; otherwise a slightly more general,
albeit still linear, form for $\bbox G$ is obtained. The factor of $1/V$ is
present to make $\lambda$ dimensionless.
The constant $\bbox G_0$ drops from the integral, as does the term proportional
to $\bbox x$, and we are left with the final result:
\begin{equation}
\label{eq:ezero-final}
\bbox E_0 = -{\lambda\over V} \int\limits_V d^3 u\, \bbox u
\left( \rho(\bbox u) - {1\over V}
  \int\limits_V d^3 v \, \rho(\bbox v) \right).
\end{equation}
In particular, for a collection of charges $\{q_{\alpha}\}$ at
positions $\{\bbox u_{\alpha}\}$, this expression becomes
\begin{equation}
\label{eq:efield-charges}
\bbox E_0 = -{\lambda\over V}\sum_{\alpha} q_{\alpha} \left(\bbox u_{\alpha} -
  \bbox u_0 \right),
\end{equation}
where $\bbox u_0$ is the center of the unit cell, given by
\begin{equation}
\bbox u_0 = {1\over V} \int\limits_V d^3 u\, \bbox u.
\end{equation}
(Notice that if $\sum_{\alpha} q_{\alpha} = 0$ the sum in
(\ref{eq:efield-charges}) is independent of $\bbox u_0$.)
Thus $\bbox E_0$ is proportional to the dipole moment of the unit cell with
respect to its center. The only remaining
freedom lies in the choice of $\lambda$ which, as shown by de Leeuw et al., is
strongly dependent on the regularization procedure.

Even though $\bbox E_0$ cannot be written as the gradient of a periodic
function we can write $\bbox E_0 = -\nabla \phi_{\text{ext}}$, where the
``extrinsic'' potential is given by
\begin{equation}
\phi_{\text{ext}}(\bbox x) = {\lambda\over V}
  \sum_{\alpha} q_{\alpha}
  (\bbox x - \bbox u_0)\mathbin{\cdot}(\bbox u_{\alpha} - \bbox u_0).
\end{equation}
Under the condition that $\sum_{\alpha} q_{\alpha} = 0$, $\phi_{\text{ext}}$
can be rewritten, up to an $\bbox x$-independent term, in the form
\begin{equation}
\phi_{\text{ext}}(\bbox x) =
 \label{eq:extrinsic-potential}
 {\lambda\over 2V} \sum_{\alpha} q_{\alpha} \,
   \vert\!\vert \bbox x - \bbox u_{\alpha} \vert\!\vert^2.
\end{equation}
Interestingly, written in this way it takes the same form as the quadratic
term in the expansion of the Ewald potential around the source:
\widetext
\begin{equation}
  \label{eq:quadratic-expansion}
  {\Phi_{_{\text{Ew}}}}(\bbox x - \bbox u) = c +
  {1\over\displaystyle\vert\!\vert{\bbox x - \bbox u}\vert\!\vert} +
  {2\pi\over 3V} \vert\!\vert{\bbox x - \bbox u}\vert\!\vert ^2 +
  \hbox{higher-order terms}.
\end{equation}
\narrowtext
This observation can be used to explain the results recently reported by
Roberts and Schnitker \cite{RobertsSchnitker:94a} and
will be the subject of a future publication.

It is important to realize that equation (\ref{eq:ezero-final}) only makes
sense after a unit cell has been chosen, which is then replicated ad
infinitum: different choices for the unit cell will give different total
dipole moments.  Fig.~\ref{fig:balls} shows schematically how the choice of
unit cell in a periodic system influences the dipole moment. The white circles
represent positively charged atoms and the black circles correspond to the
negative charges. As the figure shows, one can select different unit cells
where the white circles are on different sides with respect to the black
circles, therefore changing the dipole moment with respect to the center of
the unit cell.  Thus, if $\lambda \ne0$ the system ``remembers'' the unit cell
from which it was constructed by infinite replication.  This memory of the
unit cell complicates the dynamics and it is not clear to us what the correct
treatment is.

We should also stress that our derivation of equation (\ref{eq:ezero-final}) is
independent of any regularization (as long as the regularization respects the
symmetries of the unit cell, see above): only the value of $\lambda$ can be
regularization dependent, as shown in some specific examples by de Leeuw et
al.\ \cite{PRSL:LPS:I}.
\fi

\section{}
\label{sec:free-energies}

In this Appendix we derive some identities for the derivatives of the free
energy of a system of charges with respect to those charges.

Consider a system of $N$ fixed charges, $\{q_{\alpha}, \bbox x_{\alpha}\}$,
immersed in a solvent, constrained to a periodic box of linear dimension
$L$. We do not need to assume that the fixed
charges are pointlike. Instead, the interaction with the solvent is taken to
consist of an electrostatic piece plus a van der Waals-like short-range
repulsion.
The partition function for the system
can be decomposed into the product of a solvent-independent piece and the
``reduced'' partition function
\widetext
\begin{equation}
\label{eq:partition-function}
Z \equiv Z\left(\{q_{\alpha}, \bbox x_{\alpha}\}\right) =
 \int \prod_{j} du_j\,
 \exp\left(-\beta \sum_{\alpha,j} V_{\alpha, j}(\bbox x_{\alpha},\bbox u_j)
           -\beta H(\{\bbox u\}) \right)
\end{equation}
\narrowtext
\noindent
where the sum over $\alpha$ is over the $N$ fixed charges and that over $j$ is
over the solvent atoms. The potential $V_{\alpha,j}$ gives the
interaction energy between the charge $q_{\alpha}$ and the $j$th solvent atom,
which we take to have the form
\begin{equation}
V_{\alpha,j}(\bbox x_{\alpha}, \bbox u_j) =
 q_{\alpha} q_j \, \Phi(\bbox x_{\alpha}-\bbox u_j) +
 \Psi_{\alpha,j}^{\text{s.r.}}(\bbox x_{\alpha}-\bbox u_j)
\end{equation}
where $\Psi^{\text{s.r.}}$ is the short-range (van der Waals) interaction
and $\Phi$ is the electrostatic potential of a unit charge. For the sake of
notational simplicity we define
\begin{equation}
{\phi_{\text{solv}}}(\bbox x) = \sum_j q_j \Phi(\bbox x - \bbox u_j).
\end{equation}
The Hamiltonian $H$ contains all the energy terms that depend only on the
solvent degrees of freedom.

The aim of molecular mechanics simulations is to compute one or
several averages with respect to the Boltzmann probability distribution giving
rise to (\ref{eq:partition-function}). In this Appendix we will be concerned
with
averages that can be obtained as derivatives of $Z$ with respect to the
charges. One can also consider, of course, derivatives with respect to the
coordinates of the fixed charges (see \cite{bader:chandler} and
\cite{delbuono:94a}).

Taking a logarithmic derivative of $Z$ with respect to $q_{\alpha}$ we obtain
\widetext
\begin{eqnarray}
\label{eq:first-deriv}
{\partial\ln Z\over\partial q_{\alpha}} &=&
  -\beta \ {\displaystyle \int \prod_{j} du_j\,
 \exp\left(-\beta \sum_{\alpha,j} V_{\alpha, j}(\bbox x_{\alpha},\bbox u_j)
           -\beta H(\{\bbox u\}) \right)\,{\phi_{\text{solv}}}(\bbox
           x_{\alpha})\over
 \displaystyle
 \int \prod_{j} du_j\,
 \exp\left(-\beta \sum_{\alpha,j} V_{\alpha, j}(\bbox x_{\alpha},\bbox u_j)
           -\beta H(\{\bbox u\}) \right)} \nonumber \\
 &=& -\beta\, \langle {\phi_{\text{solv}}}(\bbox x_{\alpha})\rangle_{q}
\end{eqnarray}
\narrowtext
\noindent
where the subscript $q$ is meant to indicate that this mean value is to be
computed in the presence of the $N$ charges.

{}From equation~(\ref{eq:first-deriv}) and the relation $ \ln Z = -\beta F$
between the free energy and the partition function it follows that the free
energy of charging one ion immersed in the solvent is given by
\begin{equation}
\label{eq:free-energy-of-charging}
\Delta F _{q_i \to q_f} = \int\limits_{q_i}^{q_f} dq\,
  \langle {\phi_{\text{solv}}}(\bbox x_{\alpha})\rangle_{q} ,
\end{equation}
with a simple extension to the case of several ions in solution.
If the response to a charging process is linear \cite{gaussian},
that is, electrostriction and dielectric saturation effects are negligible,
then the mean values $\langle {\phi_{\text{solv}}}(\bbox
x_{\alpha})\rangle_{q}$ are linear in $q$ and therefore the free energy change
can be written simply in terms of the mean values at the endpoints
\begin{equation}
\Delta F _{q_i \to q_f} = {\displaystyle{q_f-q_i}\over2} \left(
  \langle {\phi_{\text{solv}}}(\bbox x_{\alpha})\rangle_{q_f} +
  \langle {\phi_{\text{solv}}}(\bbox x_{\alpha})\rangle_{q_i} \right) .
\end{equation}

The second derivatives of the free energy have also a simple form:
\widetext
\begin{equation}
\label{eq:second-deriv}
{\displaystyle\partial^2\ln Z\over\displaystyle \partial q_{\alpha} \partial
  q_{\beta}} = -\beta {\displaystyle \partial \langle
  {\phi_{\text{solv}}}(\bbox x_{\beta})
   \rangle_{q} \over\displaystyle \partial q_{\alpha}}
  = \beta^2 \biggl(\langle {\phi_{\text{solv}}}(\bbox x_{\alpha})
  {\phi_{\text{solv}}}(\bbox x_{\beta})\rangle_{q} \ -\
    \langle {\phi_{\text{solv}}}(\bbox x_{\alpha})\rangle_{q}
    \langle {\phi_{\text{solv}}}(\bbox x_{\beta})\rangle_{q} \biggr).
\end{equation}
\narrowtext
\noindent
The higher derivatives of $\ln Z$ with respect to the charges can be treated
along the same lines but they are not needed in what follows.

Note that in equation (\ref{eq:second-deriv}) there is a hidden dependence on
the positions and values of all charges, not just those indexed by $\alpha$
and $\beta$, because of the $V_{\lambda,j}$ terms in the Boltzmann weights.
This dependence is expected to disappear when $\bbox x_{\alpha}$ and $\bbox
x_{\beta}$ are both far from all the other charges and from each other. In
this regime the correlation function in equation (\ref{eq:second-deriv}) should
be independent of the value of the charges as well as the fact that they are
not pointlike. Thus in this limit this correlation function can be computed by
setting $V_{\lambda,j}$ to zero, that is, in the presence of the pure solvent.
Since these now are independent of the details of the system and only depend
on the solvent degrees of freedom, which are integrated out, we can
write down some general relations.

Let's denote by $\langle\cdots\rangle_{\text{nc}}$ the mean values
obtained by dropping these terms, i.e.,
\begin{equation}
\langle A\rangle_{\text{nc}} =
{\displaystyle \int \prod_{j} du_j\, A\,
 \exp\left(-\beta H(\{\bbox u\}) \right)\over
 \displaystyle
 \int \prod_{j} du_j\,
 \exp\left(-\beta H(\{\bbox u\}) \right)}.
\end{equation}
Since the exponents do not depend on the positions of the fixed charges,
the order of operation of derivatives, integrals and Boltzmann averaging can
be interchanged:
\begin{eqnarray}
{\partial\over\partial x_j} \langle {\phi_{\text{solv}}}(\bbox
   x)\rangle_{\text{nc}}
  &=& \langle {\partial{\phi_{\text{solv}}}(\bbox x)\over\partial x_j}
        \rangle_{\text{nc}} \\
\int d^3 x \, \langle {\phi_{\text{solv}}}(\bbox x)\rangle_{\text{nc}}
  &=&
    \langle \int d^3x \, {\phi_{\text{solv}}}(\bbox x)\rangle_{\text{nc}}.
\end{eqnarray}
In the same way we obtain the relation
\widetext
\begin{eqnarray}
\int d^3x\, &&\biggl(
  \langle {\phi_{\text{solv}}}(\bbox 0) {\phi_{\text{solv}}}(\bbox x)
  \rangle_{\text{nc}} -
  \langle{\phi_{\text{solv}}}(\bbox 0)\rangle_{\text{nc}}
  \langle{\phi_{\text{solv}}}(\bbox x)\rangle_{\text{nc}}\biggr) = \nonumber \\
 && \qquad\qquad
 \langle {\phi_{\text{solv}}}(\bbox 0) \int d^3x\, {\phi_{\text{solv}}}(\bbox
   x) \rangle_{\text{nc}} -
  \langle{\phi_{\text{solv}}}(\bbox 0)\rangle_{\text{nc}}
  \langle\int d^3x \, {\phi_{\text{solv}}}(\bbox x)\rangle_{\text{nc}}.
\end{eqnarray}
\narrowtext
Since
\begin{eqnarray}
\int\limits_V d^3x\, {\phi_{\text{solv}}}(\bbox x) &=&
  \sum_j q_j \int\limits_V d^3x\, \Phi(\bbox x - \bbox u_j) \nonumber \\
 &=& \left(\sum_j q_j\right)
     \int\limits_V d^3x \, \Phi(\bbox x)
\end{eqnarray}
is independent of the position of the solvent atoms it follows that
\begin{eqnarray}
\langle {\phi_{\text{solv}}}(\bbox 0) \int d^3x\, {\phi_{\text{solv}}}(\bbox
x)\rangle_{\text{nc}} &=&
  \\
 &{}& \kern-2em
 \langle{\phi_{\text{solv}}}(\bbox 0)\rangle_{\text{nc}}
 \langle\int\limits_V d^3x\, {\phi_{\text{solv}}}(\bbox x)\rangle_{\text{nc}}.
  \nonumber
\end{eqnarray}
{}From this it follows that
\widetext
\begin{equation}
\label{eq:no-zero-mode}
\int\limits_V d^3x\, \biggl(
  \langle{\phi_{\text{solv}}}(\bbox 0) {\phi_{\text{solv}}}(\bbox x)
  \rangle_{\text{nc}} -
  \langle{\phi_{\text{solv}}}(\bbox 0)\rangle_{\text{nc}}
  \langle{\phi_{\text{solv}}}(\bbox x)\rangle_{\text{nc}}\biggr) = 0.
\end{equation}
\narrowtext
\noindent The mean values
\begin{equation}
\langle {\phi_{\text{solv}}}(\bbox x_1){\phi_{\text{solv}}}(\bbox x_2) \cdots
{\phi_{\text{solv}}}(\bbox x_N)\rangle_{\text{nc}}
\end{equation}
are independent of any details of the fixed charges, since they are computed
in the presence of the pure solvent.

\newpage
\widetext

\end{document}